# Digital pathology-based study of cell- and tissue-level morphologic features in serous borderline ovarian tumor and high-grade serous ovarian cancer

Jun Jiang[1]*,  Burak Tekin[2]*, Ruifeng Guo[2], Hongfang Liu[1], Yajue Huang[2]#, Chen Wang[3]#.
* The two authors contribute equally to this work.
# contact authors
1 Division of Digital Health Sciences, 2 Department of Laboratory Medicine and Pathology,
3 Department of Health Science Research

**Abstract**

**Motivation:** Serous borderline ovarian tumor (SBOT) and high-grade serous ovarian cancer (HGSOC) are two distinct subtypes of epithelial ovarian tumors, with markedly different biologic background, behavior, prognosis, and treatment. According to histopathological features, several criteria are commonly used to differentiate SBOT from HGSOC, such as absence of stromal invasion or nuclear size and shape variation. However, the histologic diagnosis of serous ovarian tumors can be subjectively variable and labor-intensive as multiple tumor slides/blocks need to be thoroughly examined to search for these features. In this study, we aimed to evaluate technical feasibility of using digital pathological approaches to facilitate objective and scalable diagnosis screening for SBOT and HGSOC.

**Methods:**  We collected 15 HGSOC and 15 SBOT cases with hematoxylin and eosin-stained slides digitally scanned with 0.25μm resolution. Two pathologists assisted with manual annotations of tumor cells and stroma cells. Based on Groovy scripts and QuPath, a novel informatics system was developed to facilitate interactive annotation and imaging data exchange for machine learning purposes. Through this developed system, cellular boundaries were detected using watershed segmentation, expanded set of cellular features according to cellular components were extracted to facilitate cell-type classification using  Support Vector Machine (SVM), and patch-level histology classification of HGSOC v. SBOT was done according to Lasso-based classifiers.

**Results**: Cell-level classification was accurately achieved for both tumor and stroma cells with greater than 90% accuracy. Upon further re-examinations, 44.2% of the misclassified cells were due to over-/under-segmentations or low-quality of imaging areas. For a total number of 6,485 imaging patches with sufficient tumor and stroma cells (ten of each at least), we achieved 91-95% accuracy to differentiate HGSOC v. SBOT. When all the patches were considered for a WSI to make consensus prediction, 97% accuracy was achieved for accurately classifying all patients. A single HGSOC case was misclassified as SBOT; the review of this case revealed a spectrum of cells with remarkable size and shape variation, including a subset displaying hyperchromatic nuclei, prominent nucleoli, and high nuclear-cytoplasmic ratio.

**Conclusion**: Cellular features digitally extracted from pathological images can be used for cell classification and SBOT v. HGSOC differentiation. Introducing digital pathology into ovarian cancer research could be beneficial to discover potential clinical implications.  Larger cohort is required to consolidate our results.

1. Introduction

Ovarian cancer is one of the leading causes of death in patients with gynecological malignancies. Epithelial ovarian tumors are classified according to the cell type, with the serous type being the most common, and further divided into benign, borderline, and malignant (carcinoma) categories. Serous borderline ovarian tumors (SBOT) represent approximately 5–10% of all ovarian serous tumors [1]. Compared to benign ovarian tumors, borderline ovarian tumors exhibit greater epithelial proliferation and cellular atypia. However, in contrast to their malignant counterparts, borderline ovarian tumors lack destructive stromal invasion [2]. SBOTs share genetic changes with low-grade serous ovarian carcinomas (LGSOCs), e.g., *KRAS* and *BRAF* mutations, and can progress to the latter in a subset of patients. Overall, SBOTs are associated with a favorable prognosis, with 5-year survival rates of early-stage patients as high as 90% [1]. High-grade serous ovarian carcinoma (HGSOC), on the other hand, is a biologically distinct entity more commonly related to *TP53* mutations, most often presenting at an advanced stage and associated with a relatively poor prognosis [3]. Thus, the histologic distinction between SBOTs and HGSOCs is important in that it has major prognostic and therapeutic implications [4]. The histologic diagnosis of HGSOCs heavily relies on the morphologic assessment of tumor cells, and the presence of marked variation in nuclear size and shape represents an important feature pointing to the diagnosis of HGSOC. In general, SBOTs can be differentiated from LGSOCs and HGSOCs by lack of destructive stromal invasion and high-grade morphologic features, respectively. However, SBOTs can display "micro-invasion" and high-grade morphologic features may be very focal in a given HGSOC case, potentially posing diagnostic challenges and necessitating extensive evaluation of multiple surgical blocks of resected specimens [3]. Overall, the accurate histologic diagnosis of serous ovarian tumors can be labor-intensive and subjective.

With the advancement of digital pathology (dPath), there have been substantial research interests in developing image analysis approaches for digitally characterizing tissue heterogeneity in various cancer types and studying associations with clinical outcomes. For example, Lu et al.[5] found that the nuclear shape and orientation features are predictive of survival of early-stage estrogen receptor-positive breast cancers. Lan et al. [6] showed that quantitative measurements of the extent and density of lymphocytic infiltration was significantly associated with overall survival and progression-free survival in ovarian cancers. Beck et al.[7] showed that nuclear morphologic features within the stroma were significantly associated with survival in breast cancer. Sidra et al. [8]revealed that tumor spatial heterogeneity was a strong prognostic factor in ovarian cancers. Moreover, dPath approaches were also developed for facilitating diagnoses of different tumor subtypes within a single cancer type, e.g. Zhang et al.[9]proposed a stepwise method to classify Non-Hodgkin's lymphoma subtypes, over 99% cross validation accuracy was achieved on differentiating chronic lymphocytic leukemia, follicular lymphoma, and mantle cell lymphoma. Rathore et al. [10] demonstrated that deep learning techniques are capable of predicting overall survival and molecular markers (isocitrate dehydrogenase gene mutation and co-deletion of chromosomes 1p and 19q) in gliomas. Barker et al. [11]proposed a brain tumor subtype classification model; spatially localized features of shape, color and texture from whole slide image tiles were eligible in differentiating glioblastoma multiforme and lower grade glioma.

Although SBOT and HGSOC taxonomically belong to the same umbrella category of ovarian epithelial tumors, they are markedly distinct entities with dissimilar biologic behavior and histologic findings. Based on the latter, it can be anticipated that they have different cellular composition and spatial heterogeneity. Motivated by the aforementioned diagnostic challenges

(subjective and labor-intensive histopathologic examinations), we sought to investigate dPath feasibility of classifying HGSOC and SBOT from cellular and tissue levels using machine learning (ML) approaches. The designing principals of proposed dPath-ML approaches include interactive pathological annotations, objective quantifications of imaging features, and scalable computations to thoroughly scan entire WSI. The overall flow of this developed informatics process is shown in **Fig. 1**: it started from interactive pathology-annotation and machine-abstraction through QuPath-Groovy data communication. Then, cellular features were extracted to train ML predictive models to classify cell types, and tissue-level histology classifications were done by aggregating cellular level features. Overall, we achieved accurate prediction results from cellular and tissue levels, strongly suggesting great potentials for applying dPath methods to broad research and clinical applications.

## 2. Materials and Methods

### Cohort selection

Thirty cases (15 cases of SBOT and 15 cases of HGSOC) with unequivocal histologic features were randomly retrieved from the SoftPath system database. The diagnoses were independently confirmed by two pathologists. Clinical features of the cases are outlined in Table 1. The Regional Institutional Review Board approved the study (IRB ID: 19-011615).

### Image Acquisition and Annotation

Archived slides were delivered to Pathological Research Core (PRC) in Mayo Clinic, which provides advanced histology-related services, including immunohistochemistry, tissue microarray construction, and digital imaging and analysis for research and clinical investigators at an affordable cost with reasonable turnaround times. The magnification of digital whole-slide scanner (Aperio Scanscope XT) was set to 40x, with 0.25μm*0.25μm of pixel size in obtained whole slide images (WSIs).

For ten WSIs as training-set (5 each for SBOT and HGSOC), two pathologists assisted for annotations. In each case, five regions most representative of SBOT/HGSOC morphology features were chosen as regions of interest (ROI) for manual annotations, with each ROI being larger than 256μm x 256μm. Cells in the ROIs were annotated to be tumor or stroma cells. Rather than directly delineating the boundary of each cell and then assigning labels to each cell, we proposed a simplified annotation process, in which pathologists were invited to use polygons to contour homogeneous regions where cells were considered to share the same label, except those deliberately annotated by points. Based on the built-in interface in QuPath, we made customized scripts to process pathologists' annotations (details seen in section below).

**Groovy-based interactive pathology-annotation and data abstraction**
As a powerful tool for quantitative pathology and bio-image analysis, QuPath provides application programming interface (API) to enable high-throughput manner across many images, which greatly extends the feasibility of customizing pathological image processing pipelines. Groovy was officially elected by QuPath group as the particular programming language to communicate with QuPath for image processing, interactive annotation and visualization, since it closely matches the Java programming language in which the majority of QuPath itself was written. In order to build a dPath framework with flexible pathology-annotations and expandable

ML modules, we designed Groovy-based interaction middleware to communicate between QuPath and Python (or any programming languages with ML modules), which played important roles in our framework.

1) Annotation module (**Fig. 2** ①). The main object of this module is to reduce pathologists' annotation workloads from circulating the boundary of cells. Homogeneous regions and points annotated by pathologists were processed with the aid of customized scripts and built-in interface in QuPath. Within this process, cells were detected with watershed segmentation plugin, so that category labels from annotations (homogeneous regions and points) could be assigned to individual cells. Some parameters of this cell segmentation algorithm were fine-tuned according to our dataset: pixel size was set to 0.25 $\mu$m, minimum nuclear area was set to 10 $\mu m_2$

2) Cellular feature extraction module (**Fig. 2** ②). After cell segmentations, cellular features were extracted from SBOT and HGSOC cases, according to morphological, color-intensity characteristics. These features were constructed according to different cellular components, i.e. nuclear, cytoplasm and cell: Nuclei were automatically segmented from the background using a watershed nuclear segmentation method. The boundaries of nuclei were arbitrarily expanded up to 5 $\mu$m or until the expansions overlapped with adjacent cells. The extended areas were regarded as cytoplasm. Cells were the integration of nuclear and cytoplasm. Morphological features were calculated based on the binary mask of each nucleus/cytoplasm/cell, which were used to describe the geometry properties of the cells, including area, perimeter etc. In order to enrich color-intensity characteristics of nucleus/cytoplasm/cell, original H&E images were decoupled into Hematoxylin and Eosin components. Color-intensity features were calculated based on the image attributes under the binary mask of each cell, including Hematoxylin and Eosin optical density (OD) mean, standard deviation, etc. With customized scripts, all the features were calculated in QuPath, and exported into csv files, in which each row is a sample of a cell, the first column is the annotated label of the cell, and the remaining columns are feature values.

3) Visualization module (**Fig. 2** ③). Cell/tissue classification and some intermedia results (such as binary images, polygon coordinates etc.) were converted into QuPath objects and imported into QuPath for convenient result visualization and examination.

In a nutshell, Groovy will export pathologist's manual annotations, such as cellular coordinates in images and cell-types labels, into ML module for building predictive models; once a predictive model is built, cellular and tissue-level results can be returned by ML module and displayed in QuPath through coordination of Groovy communication module. This enables further examinations of misclassified cells and tissue regions, for both quality assurances and iterative improvement of model-training processes. All the code and relevant documentations are available to the public in our GitHub (https://github.com/smujiang/CellularComposition).

# Image analysis
**Cellular Classification and Examinations of feature importance**
With N=41 extracted features, linear SVM (Support Vector Machine) was introduced to classify cells and clarify the importance of features in cell type identification. As a state-of-the-art ML technique, each cell in SVM was interpreted as a sample in the feature space, and the algorithm was trained to fit a hyperplane (decision boundary) which tries to maximize the margin between sample types. Support vectors were samples (cells) to determine the decision boundary.

Distances from sample to hyperplane were abstracted to reflect the likelihood of a sample to be correctly classified. Once the hyperplane was determined, the coefficients of the trained model were used to determine feature importance of cell classification, since the weights figure the orthogonal vector coordinates orthogonal to the hyperplane and their direction represents the predicted class. Line chart was used to show the difference in feature importance for cell classification tasks in SBOT and HGSOC respectively. Confusion matrix was calculated to show the cell classification accuracy.

**Tissue Patch Classification for SBOT v. HGSOC Differentiation**
For each SBOT and HGSOC (15 vs. 15) case, 10 ROIs with both tumor and stroma areas were selected for classification evaluation purposes. Cell classification model trained in previous phase was applied to these ROIs to differentiate tumor and stroma cells. ROIs were divided into regular 512 x 512 pixel patches to enable measurements of local differences. Patches that contain at least 10 tumors and 10 stroma cells were considered to carry enough information to differentiate SBOT v. HGSOC tissues. For each eligible patch, total n=609 features were aggregated from cell-level results, including 1) statistics of cellular features. Starting from cellular features, we aggregated each feature to patch-level considering overall distributions: mean, median, standard-deviations, [Q1, Q3], minimum and maximum values per cell-type. 2) Tumor-stroma interaction features: As tumor cells tend to aggregate to clusters, we used Gaussian-based kernel density estimation (KDE) to fit empirical density distribution of tumor cells within a patch. Then, this fitted KDE function was used to evaluate relative distance of stromal cells with respect to tumor cell clusters. In order to capture tumor-stroma interaction from multiple scales, KDE kernel width was set to 16, 20, 24, 30 and 34. For each stromal cell, we calculated probability/likelihood score based on KDE, and 7 statistics of the scores were included in our patch descriptor.

Considering the high dimensionality of patch features, LASSO regression coefficients were used to highlight the important features with most critical contributions to patch classification performance. In LASSO cost function, the penalty term regularizes the coefficients such that the coefficients that take large values get penalized, resulting in shrinking the count of non-zero coefficients which helps to reduce the model complexity and multi-collinearity.

## 3. Results

**Cell Classification**
With 10 (5 v. 5 SBOT, HGSOC) training WSIs, we annotated 17,181 tumor cells and 8,828 stroma cells in HGSOC cases (on average 3,436 tumor cells and 1,766 stromal cells per WSI), and 2,638 tumor cells and 6,435 stroma cells in SBOT cases (on average 527 tumor cells and 1,287 stromal cells per WSI). Using SVM classifier based on 41 cellular features, we achieved 86.4%-89.1% cell classification accuracy in HGSOC, and 85.4% - 90.8% in SBOT cases (Fig. 3 A & B) with leading features shown as Fig. 3.C; among them, Eosin OD intensity was found to play the leading role in differentiating stroma and tumor cells in both categories. (**Fig. 3 C**). We also conducted unsupervised clustering to investigate features' inter-correlation, showing that features with similar morphology and intensity implications were often highly correlated (**Supplementary Fig. 1**).

In order to shed some lights on cell-level misclassifications, KNN algorithm (K=10) was used to cluster 3,575 misclassified cells. Through manual examinations, 44.2% (1,580) of them could be attributed to explainable process errors, such as under/over cellular segmentation, histologic artifacts induced by slide preparation/scanning, or non-specific cellular/tissue elements that could not be unequivocally identified as either tumor or stroma cells (e.g., portions of adipocytes or air/fat bubbles, heavily pigment-laden cells, red blood cells/hemorrhage, possible non-cellular connective tissue fragments) (**supplementary Fig. 2**). Besides of examining individually misclassified cells, we also investigated whether some support vectors, as representative cells for SVM classifier, may contribute to systematic classification errors. Similar to the misclassified cells, the classification of cells associated with support vectors was also complicated by the presence of the aforementioned nonspecific elements, which indicates that elaborative features for these cells are essential for performance improvement. Of note, when taken out of their context, a subset of misclassified cells was challenging to confidently classify as tumor or stroma cells for the pathologists, which indicates potential morphologic variability between different cases and reflects the pathologists' practice of evaluating the cells in light of the totality of the slide in a given case.

**Tissue- and subject-level Differentiations of SBOT v. HGSOC**
In order to conduct tissue- and subject-level classifications, 300 ROIs (150 v. 150 HGSOC, SBOT) were selected with cell detection and classification done as previously described (section Image analysis). We detected 903,678 cells from HGSOC ROIs, of which 404,973 were tumor cells; we detected 465,299 cells from SBOT ROIs, of which 151,393 were tumor cells. For evaluating tissue-level discrimination between SBOT and HGSOC, we divided ROIs to multiple regular image patches of 512 x 512 (Fig. 4A and 4B). In total, we obtained 6,446 image patches from HGSOC and 4,025 patches from SBOT cases. Based on cellular features and tumor-stroma spatial distributions, we summarized 609 patch-level features for each image patch. In particular, we varied kernel width of KDE to provide multi-resolution features for characterizing tumor-stroma reaction. With our rules (at least 10 tumor and 10 stroma cells in a patch), 6,485 (HGSOC: 4,225, SBOT: 2,260) image patches were selected from 10,471 (HGSOC: 6,446, SBOT: 4,025) patches as eligible for tissue (patch) classification. With 609 dimensional features, SVM was trained to differentiate SBOT and HGSOC. We achieved overall patch-level accuracy of 90.5~90.7% (Fig. 4 C). In order to evaluate classification separations, distances of image patches to SVM classification hyper-plane were computed, and distances per histotype histograms demonstrated that aggregation of cellular features has largely separate HGSOC v. SBOT (Fig. 4D). To further delineate features with most critical contributions to classification performance, sparsity-based LASSO regression was applied, which revealed 15 features with non-zero coefficients (Table 2). We found that geometry features, such as cell area and perimeter, did not play a significant role in cell classification, but were important for tissue differentiation. Statistical values from hematoxylin intensity of tumor and stroma were strongly associated with SBOT v. HGSOC classification. Interestingly, we found that our KDE features did not capture a significant pattern for SBOT v. HGSOC differentiation. Moreover, when aggregating multiple tissue patches' scores into subject-level with bootstrap resampling, 97% (29/30) accuracy was achieved (Fig. 4 E), which shows that tissue-level features were potentially significant for subject classification.
Through histopathologic reviews of misclassified patches (n=199), several morphologic features were repeatedly found in multiple patches and cases. For example, (i) some patches from

HGSOC had compact populations of cells with nearly overlapping cellular borders, which might have arguably influenced accurate segmentation of cells. (ii) Some HGSOC cases had a predominance of cells with an optically clear nuclear and/or cytoplasmic appearance. A possible explanation for the misclassification of these patches may be related to their overall lower Eosin or Hematoxylin intensity compared to other HGSOC cases with more hyperchromatic (darker) nuclei. (iii) Another common feature among misclassified patches was the presence of a spindle-cell population aligned in a streaming fashion (**supplementary Fig. 4**)

**Discussion**
Although abundant work has been done on cell segmentation and classification [refs], these two tasks are still open questions. For cell segmentation, approaches were fraught with over- and under-segmentation due to high variation of shapes and textures[12]. These limitations were also found in this work. An automatic cell segmentation method (Watershed Segmentation) was introduced into our work to exempt pathologists from directly delineating the cell borders in the annotation step. According to our experience, over- and under-segmentation were more likely to be observed in tumor cells, which may be attributable to the greater variation of both shape and texture of tumor cell nuclei in comparison to stromal cell nuclei. Of note, lymphocytes were less likely to be over- or under-segmented due to their relatively uniform morphology. The consequence of segmentation errors in automatic methods is that the error will propagate to the following steps, including morphological and texture feature extraction, cell classification model training, etc. For example, one cell may be over-segmented into several parts, while two adjacent cells without clear boundaries may be under-segmented into one cell; with both instances leading into statistical errors in cell radius.

Several features of the study methodology and findings deserve further mention from the pathologists' perspective. The cellular features evaluated herein overlap with some of the parameters that are routinely assessed in the surgical pathology practice, such as nuclear-to-cytoplasmic ratio or the presence of hyperchromatic nuclei. However, the algorithm of this study in its current form does not address two features that are frequently evaluated by pathologists when differentiating benign and malignant lesions, namely pleomorphism and detection of nucleoli. The former refers to size and shape variation of cells and the latter refers to small spherical structures in the nucleus. The presence of marked pleomorphism or prominent nucleoli may favor malignancy, within the appropriate context. **Supplementary Figure 4** shows examples of misclassified patches from HGSOC cases that, upon review, were observed to have pleomorphism and/or prominent nucleoli. The inclusion of these features into the analysis can further refine the classification accuracy in follow-up studies.

In pathologists' daily practice, the final diagnosis represents an overall and somewhat subjective interpretation of the totality of many findings, including clinical factors such as patient's age, size or growth rate of the tumor, as well as morphologic features on the slide, and ancillary studies such as immunohistochemical stains. As such, it is challenging for any mathematical model to approximate the human diagnostic thinking process. Of note, a given histologic slide may include areas of varying morphologic characteristics, which was also observed in this study. What further complicates development of mathematical models and digital pathology algorithms is that, on a case-by-case basis, certain morphologic features can potentially "trump" or "overrule" others from the pathologists' perspective. The only misclassified case in the study (Case 12) represents an example for this phenomenon in that it harbors large cells with bizarre, hyperchromatic nuclei and high nuclear-cytoplasmic ratio. A review of the misclassified patches of this case (75 misclassified out 132 eligible patches) showed that these cells were outnumbered by the surrounding, more uniform population of smaller cells, which might have skewed the algorithm towards misclassifying these patches as borderline. However, in daily practice, the presence of the large cells with *bizarre*, hyperchromatic nuclei -even if few in number- would "trump" other morphologic features and make the pathologist favor a diagnosis of HGSOC over SBOT. Another noteworthy feature in this misclassified case is the presence of prominent nucleoli in a subset of cells **(Figure 5).** Further refinement of the study algorithm to incorporate the evaluation of "outlier" cells with remarkably different, *bizarre* sizes and shapes, overall cellular pleomorphism, and prominent nucleoli would improve classification accuracy, especially in cases like Case 12.

In recent years, deep learning based approaches were employed for cell segmentation and classification in many cancer research tasks [13-15]. These advanced methods are considered to be superior to traditional ML methods, as cell classification model for one cancer could be easily retrained with transfer learning strategy for another cancer. However, morphologic features across cancers and their subtypes could be dramatically different, which means that large numbers of cells need to be annotated for specific cancers due to the data-hungry nature of these deep learning models, making cell annotation impractical for many application scenarios. Even if deep neural networks are trained for a specific task, interpretability will still be poor. In deep learning models, it is hard to explain which elements play a more important role in a downstream analysis because feature extraction components are embedded in some layers of the networks. Investigators have to use step-wise methods (detect individual cells, extract features for each cell, and classify cells according to the features) to evaluate the significance of a specific feature, and connect this information to molecular/genomic discoveries.

In clinical practice, two main histologic features are utilized to differentiate between SBOT and HGSOC, namely stromal invasion and cellular morphology. This study with its current methodology, however, does not address the stromal invasion, but mainly focuses on the evaluation of morphologic features. This represents a limitation, and we are planning future studies investigating stromal invasion. Nevertheless, this point could also be perceived as a relative strength since, solely using the cellular morphologic features, our current algorithm was able to accurately classify the vast majority of the cases.

In this work, decoupled H&E intensity information played an important role in cell classification, even if the cells were not well segmented. In practice, a limitation of employing H&E intensity for classification purposes is that this parameter depends on multiple factors, such as the histologic staining process or the duration since the initial preparation of slides as the latter typically tend to fade with time. Despite these limitations, our cell classification accuracy reached up to 87.8%. Our further work will put more efforts on developing cell segmentation and classification methods with better performance, and incorporating domain-specific histologic features such as pleomorphism and prominent nucleoli into the analysis.

**Tables.**

**Table 1**. Clinical characteristics of the study cases.

| | Case number | Age at diagnosis (yrs) | Year of diagnosis | Stage at diagnosis | Progression, recurrences, or metastases during follow-up | Duration of follow-up (mos) |
|---|---|---|---|---|---|---|
| High-grade serous ovarian cancer | 1 | 69 | 2011 | IC1 | None | 105 |
| | 2 | 51 | 2015 | IIIC | Progressive metastatic disease | 21 |
| | 3 | 64 | 2011 | IIIC | Progressive peritoneal carcinomatosis | 34 |
| | 4 | 48 | 2004 | IIIC | Clinical and biochemical disease progression | 3 |
| | 5 | 86 | 2004 | IIIC | None* | 36 |
| | 6 | 73 | 2002 | IIIC | Clinical and biochemical disease progression | 126 |
| | 7 | 77 | 2009 | IIIC | Progressive metastatic disease | 34 |
| | 8 | 47 | 2008 | IIIC | Progressive metastatic disease | 45 |
| | 9 | 76 | 2007 | IIIC | Progressive metastatic disease | 23 |
| | 10 | 69 | 2010 | IIIC | Progressive peritoneal carcinomatosis | 19 |
| | 11 | 48 | 2012 | IIIC | None | 91 |
| | 12 | 60 | 2006 | IC3 | None | 165 |
| | 13 | 86 | 2001 | IIIC | Progressive metastatic disease* | 5 |
| | 14 | 55 | 2012 | IC1 | None | 91 |
| | 15 | 71 | 2003 | IC3 | Local recurrence | 200 |
| Serous borderline ovarian tumor | 16 | 80 | 2006 | IA | None | 37 |
| | 17 | 20 | 2001 | IA | None | 224 |
| | 18 | 67 | 2009 | IA | None | 129 |
| | 19 | 71 | 2002 | IA | None | 207 |
| | 20 | 35 | 2002 | IIIC | Diagnosed with low-grade serous ovarian carcinoma 141 months after initial diagnosis of serous borderline ovarian tumor, subsequently developing lymph node metastases and disease progression | 215 |
| | 21 | 38 | 2003 | IA | None | 198 |
| | 22 | 46 | 2014 | IA | None | 76 |
| | 23 | 69 | 2005 | IIIC | None | 183 |
| | 24 | 28 | 2004 | IIIC | None | 185 |
| | 25 | 29 | 2006 | IC3 | None | 169 |
| | 26 | 60 | 2002 | IIB | None | 219 |
| | 27 | 64 | 1999 | IIC | Diagnosed with low-grade serous ovarian carcinoma 75 months after initial diagnosis of serous borderline ovarian tumor, with progressive metastatic disease | 89 |
| | 28 | 59 | 2012 | IA | None | 91 |
| | 29 | 64 | 2001 | IIIA2 | Diagnosed with low-grade serous ovarian carcinoma 171 months after initial diagnosis of serous borderline ovarian tumor, with progressive metastatic disease | 187 |
| | 30 | 46 | 2001 | IA | None | 223 |

Staging based on FIGO's staging classification for cancer of the ovary, fallopian tube, and peritoneum (2014) [16]
*Limited clinical follow-up information available.

**Table 2.** Important features selected by Lasso classifier.

| No. | Content | Cellular Feature Name | Statistics | Importance |
|---|---|---|---|---|
| 1 | stroma_Cell | Eosin OD min | min | 0.11 |
| 2 | stroma_Cell | Hematoxylin OD max | Q3 | 0.33 |
| 3 | stroma_Nucleus | Hematoxylin OD max | min | -0.33 |
| 4 | stroma_Nucleus | Hematoxylin OD max | Q1 | -0.35 |

| 5 | tumor_Cell | Area | mean | 0.1 |
| 6 | tumor_Cell | Eosin OD max | max | 0.21 |
| 7 | tumor_Cell | Eosin OD max | Q3 | 0.23 |
| 8 | tumor_Cell | Hematoxylin OD max | Q3 | 0.14 |
| 9 | tumor_Cytoplasm | Hematoxylin OD max | median | -0.28 |
| 10 | tumor_Cytoplasm | Hematoxylin OD mean | max | -0.9 |
| 11 | tumor_Cytoplasm | Hematoxylin OD max | Q3 | -0.24 |
| 12 | tumor_Cytoplasm | Eosin OD max | max | 0.26 |
| 13 | tumor_Nucleus | Perimeter | mean | -0.12 |
| 14 | tumor_Nucleus | Hematoxylin OD max | max | 0.17 |
| 15 | tumor_Nucleus | Hematoxylin OD mean | Q3 | -0.16 |

**Figure Legend**

**Figure 1**. Overall informatics framework for digital pathology analysis of SBOT vs. HGSOC differentiation.

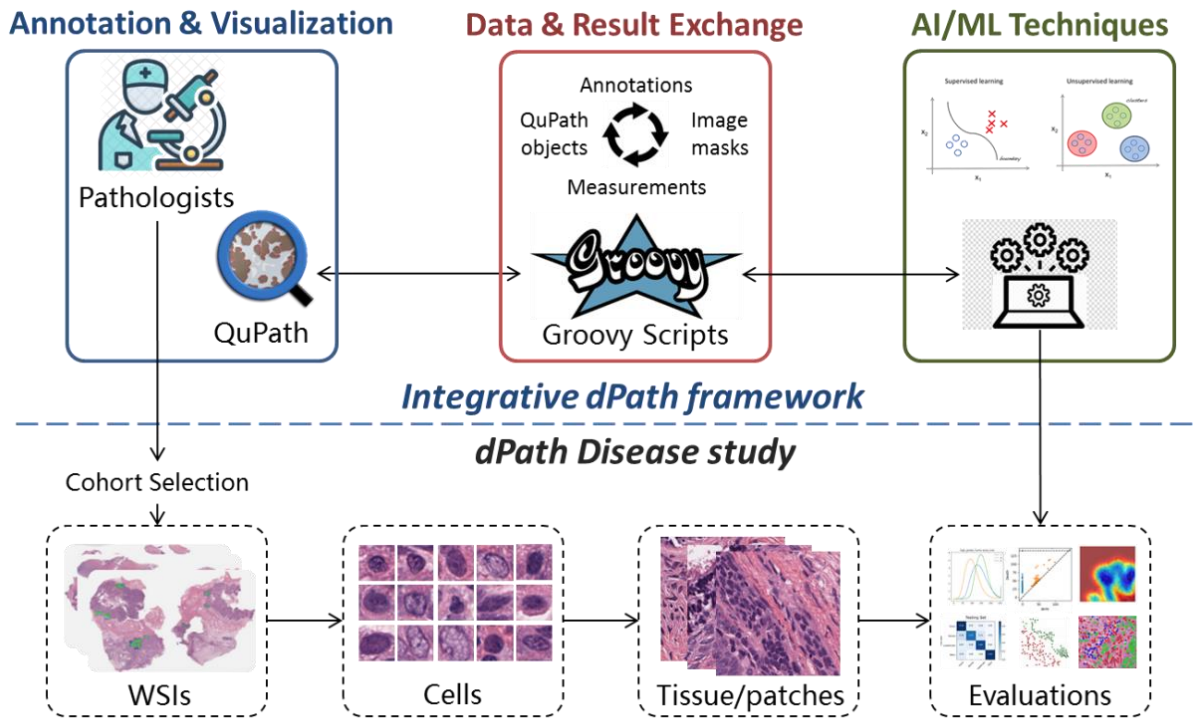

**Figure 2.** Relevant major steps for Groovy-based annotation and data abstraction. Each blue arrow denotes a Groovy middleware, including ①annotation processing, ② feature extraction and ③result visualization.

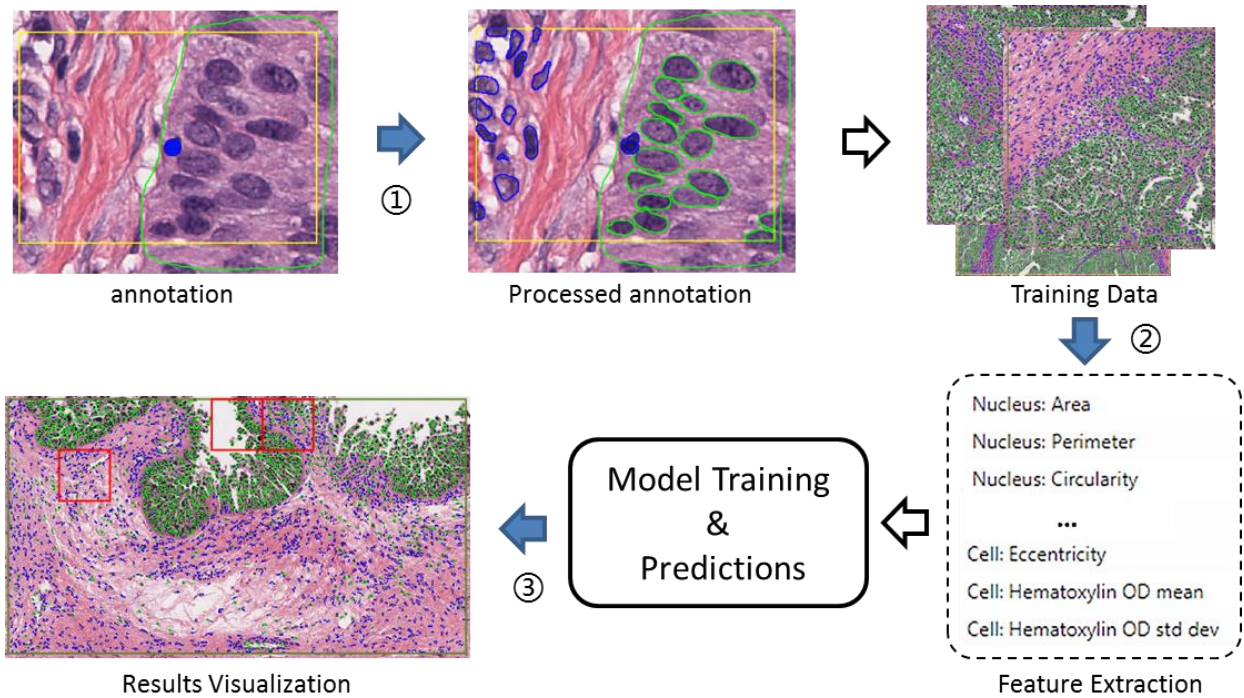

**Figure 3.** Cellular-level classification results. (A) and (B) Confusion matrix of cell classification in HGSOC and SBOT. (C) feature ranking based on SVM classifier. Feature importance for cell classification in high grade and borderline cases. Feature importance (X axis) was normalized to [-1, 1]. Ticks of Y axis are the features (sorted by the importance of HGSOC cases) selected by SVM for tumor and stroma classification.

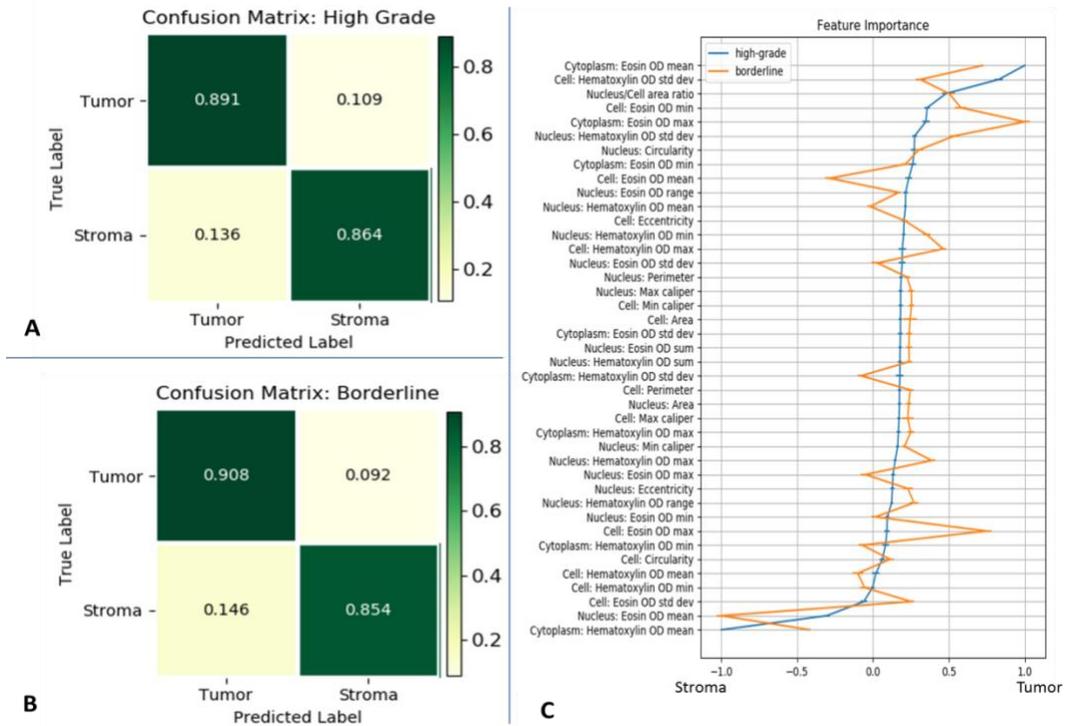

**Figure 4.** Tissue- and subject-level classification results. (A) Cell detection and classification were conducted in several ROIs per WSI. (B) ROIs with cell labels were divided into regular image patches. (C) ROC curve of patch classification with aggregated cellular features. (D) tissue-level histograms of distances to SVM hyper-plane; (E) subject-level aggregated distances after n=1000 bootstrapping.

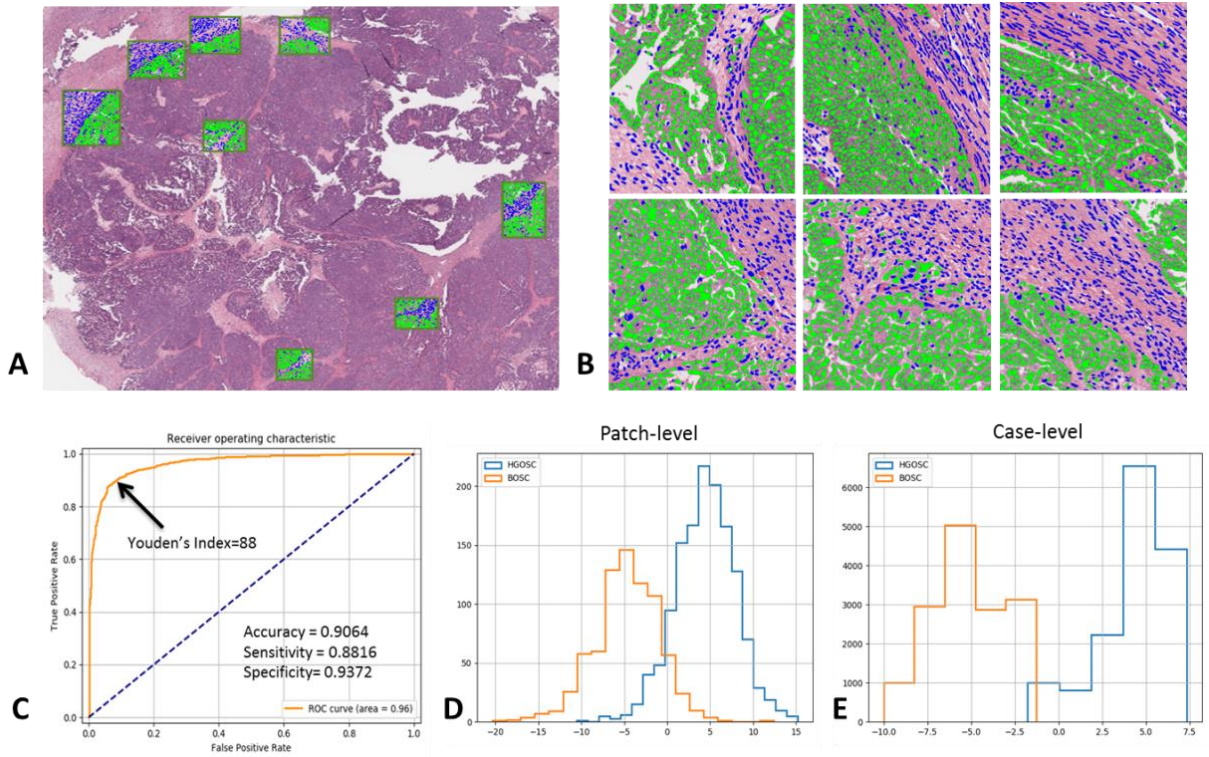

**Figure 5.** Image patches from case 12, the only misclassified case in the study. There are multiple cells with nondescript, *bizarre* shapes and hyperchromatic nuclei (red arrows). In practice, the presence of these cells, even in the absence of other worrisome features, would significantly raise the level of the pathologists' concern about a high-grade malignancy. Some of the cells in **C** and **D** appear to have two nuclei. **E** and **F**, captured from a different area of the WSI, demonstrate a strikingly different morphology, with a distinct population of cells displaying less hyperchromatic nuclei with discernible nucleoli. Some of the cells appear merged together, forming giant cells with a somewhat *syncytial* appearance (green arrows). Hematoxylin and Eosin. The image patches in each panel are 512 pixels x 512 pixels and approximately correspond to 200x magnification.

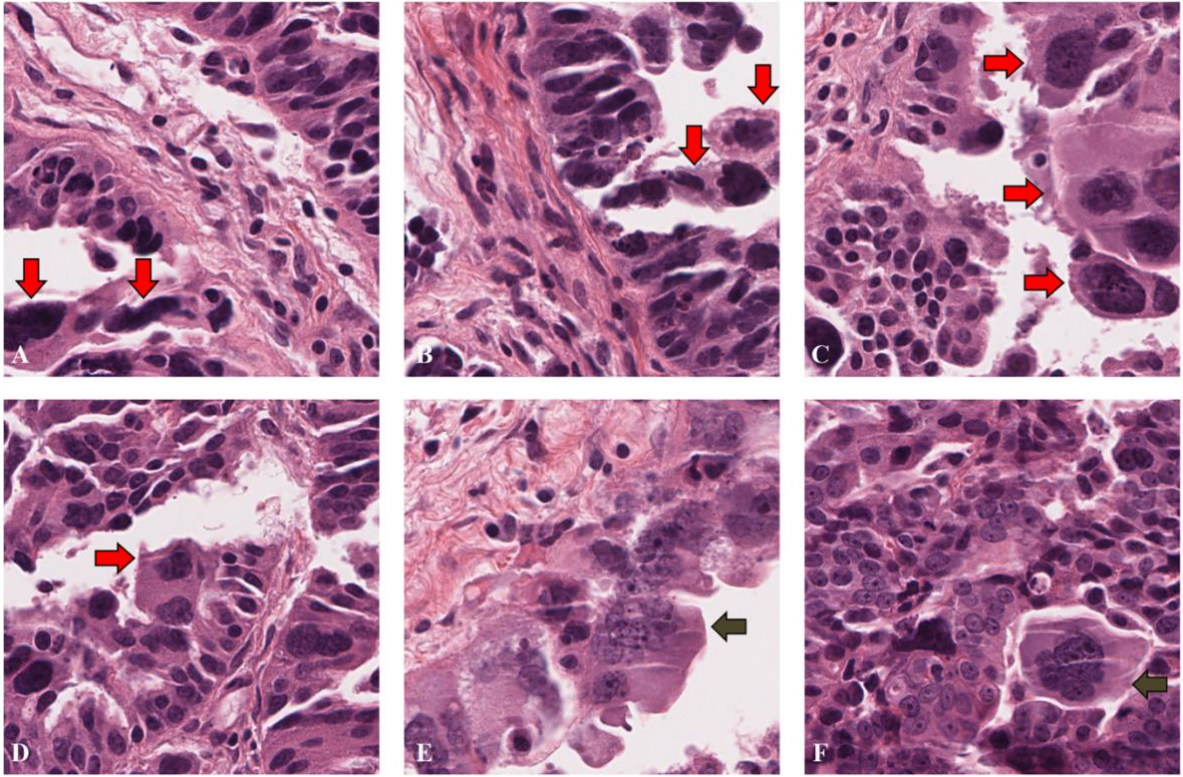

**Supplementary Figure Legends**

**Supplementary Fig. 1.** Correlation heatmap of cellular features.

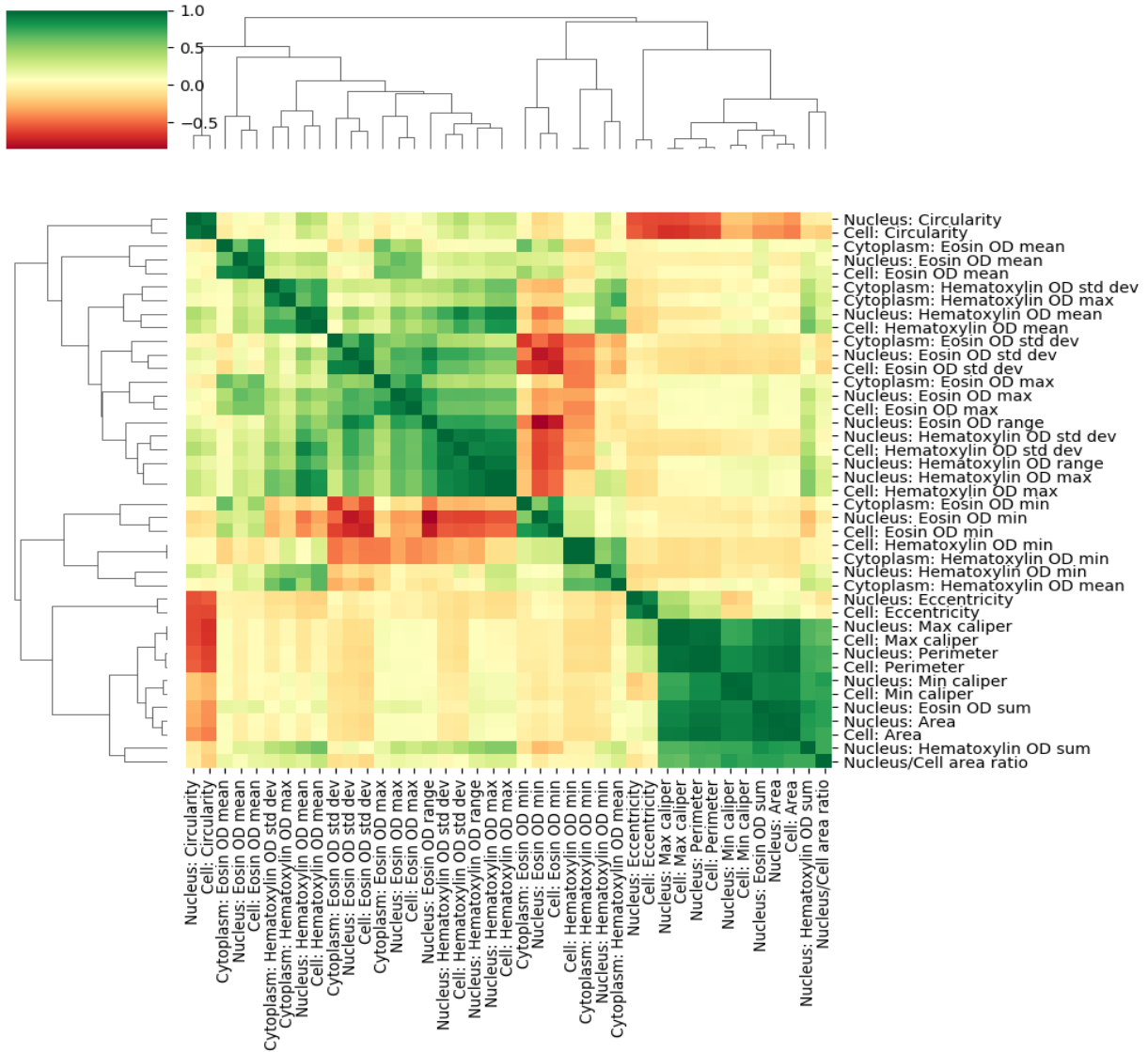

**Supplementary Fig. 2.** Examples of misclassified cells. A – C exemplifies over-segmentation. D and E: portions of adipocytes and/or bubbles of fat; F: heavily pigment-laden cells; G: red blood cells, H: plasma cell; I and J: possible non-cellular connective tissue fragments or cells with indistinct borders.

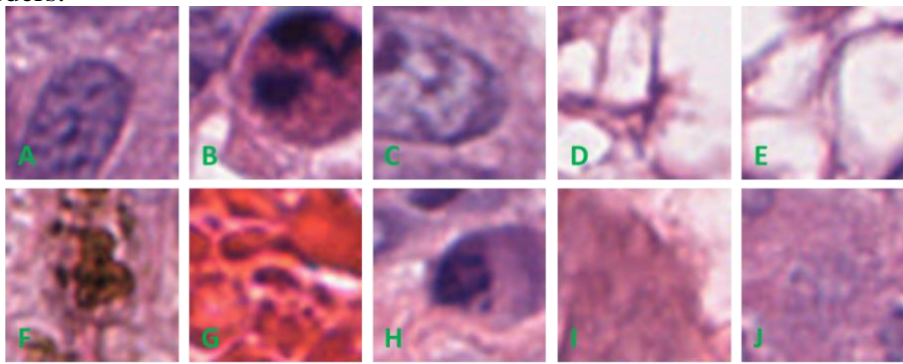

**Supplementary Fig. 3.** Examples of misclassified image patches. **A and B** (Case 5 in Table) harbor cells with indistinct or nearly overlapping cellular borders. The majority of the cells in **C and D** (Case 13) have optically clear nuclear and/or cytoplasmic appearance. Spindle cells aligned in a curvilinear or linear directionality are present in **E** (Case 9) and **F** (Case 22), respectively. Marked variation in cellular shape and size, indicated by the blue arrows (pleomorphism), can be appreciated in **G** (Case 12). Of note, the cells have hyperchromatic nuclei (see C and D for comparison). There is a mixture of cells with a spectrum of nuclear contours and chromatin profile in **H** (Case 5). Cells with prominent nucleoli (orange arrows) are seen in I and J (Cases 2 and 5, respectively). Hematoxylin and Eosin. The image patches in each panel are 512 pixels x 512 pixels and approximately correspond to 200X magnification.

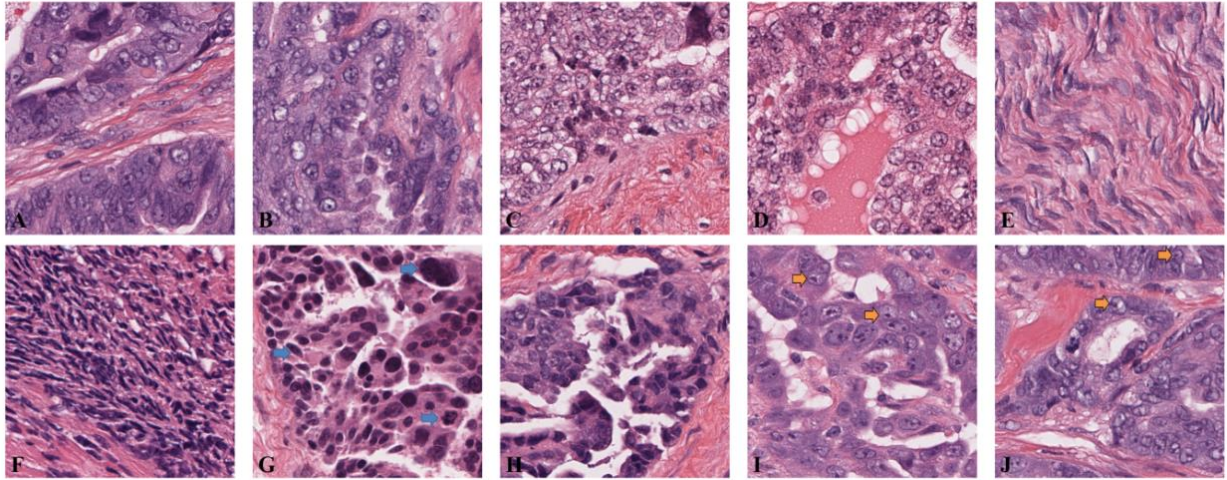